\documentstyle[preprint,prb,aps]{revtex}

\begin{document}
\draft
\title{Quantum diffusion on a cyclic one dimensional lattice}
\author{A. C. de la Torre, H. O. M\'{a}rtin, D. Goyeneche}
\address{Departamento de F\'{\i}sica,
 Universidad Nacional de Mar del Plata\\
 Funes 3350, 7600 Mar del Plata, Argentina\\
dltorre@mdp.edu.ar}
\date{\today}
\maketitle
\begin{abstract}
The quantum diffusion of a particle in an initially localized
state on a cyclic lattice with $N$ sites is studied. Diffusion
and reconstruction time are calculated. Strong differences are
found for even or odd number of sites and the limit
$N\rightarrow\infty$ is studied. The predictions of the model
could be tested with micro - and nanotechnology devices.
\end{abstract}

\section{INTRODUCTION}
The problem of a classical particle performing a random walk in
various geometrical spaces\cite{crwf} and the quantum random
walk\cite{qrw} have been thoroughly studied and compared. The
classical and quantum cases have striking differences. One of
these differences is that, whereas the classical spread increases
with time as $\sqrt{T}$, in the quantum case we have a stronger
linear time dependence of the width of the distribution.  In the
quantum mechanical case, we can identify two different causes for
the spreading of the probability distribution describing the
position of a particle. There is a spreading of the distribution
caused by the random walk itself, also present in the classical
case, and superposed to it, there is the quantum mechanical
spreading of the probability distribution due to the time
evolution of a particle in a localized state. This second type of
spreading is the main interest of this contribution. For this
study, we will consider a quantum mechanical particle initially
localized in one site of a one dimensional cyclic lattice with
$N$ points. In most treatments of quantum random walks in a
lattice it is assumed that the number of sites, $N$, is large
compared with the number of jumps of the time evolution and
therefore the system does not notice whether the lattice is
infinite or cyclic, that is, finite with periodic boundary
conditions. In our analysis we will not assume that $N$ is large
and we will find some peculiar and interesting features like, for
instance, a very different behaviour for even or odd values of
$N$. There are several motivations, besides the general academic
interest, for allowing low values of $N$. For instance, cyclic
lattices with a few sites have been built (quantum corrals) with
nanofabrication techniques and in quantum computers we deal with
systems with $N=2$ (qubits) or $N=3$ (qutrits). In all these
cases we may be interested in the quantum diffusion time of an
initially localized state. The quantum behaviour in the
continuous case $N\rightarrow\infty$ is also interesting because
it can be experimentally tested building small conducting rings
with microfabrication techniques.
\section{DEFINITION OF THE MODEL}
In this work we will consider a particle moving in a one
dimensional periodic lattice with $N$ sites and lattice constant
$a$ represented in Figure 1. The quantum mechanical
treatment\cite{dlt} of this system requires an $N$ dimensional
Hilbert space ${\cal H}$. The lattice sites will be labelled by
an index $x$ running through the values $0,1, \cdots, N-1$. We
will adopt a very useful notation for the principal $N^{th}$ root
of the identity defined by
\begin{equation}\label{ome1}
\omega =e^{i\frac{2\pi}{N}}\ .
\end{equation}
Integer powers of this quantity build a cyclic group with the
important property
\begin{equation}\label{cycl}
  1=\omega^{Nn}\ , \forall n=0,\pm 1,\pm 2,\cdots \ .
\end{equation}
The position of the particle in the lattice can take any value
(eigenvalue) $a(x-j)$ where $a$ has units of length, $j=(N-1)/2$
and the integer number $x$ can take any value in the set $\{ 0,1,
\cdots , N-1 \}$. The eigenvalues have been chosen in a way that
 position can take positive or negative values in the interval
$[-aj, aj]$. Notice that $j$ is integer for odd $N$ and
half-odd-integer if $N$ is even. The state of the particle in
each position is represented by a Hilbert space element
$\varphi_{x}$ and the set $\{\varphi_{x}\}$ is a basis in ${\cal
H}$. In the spectral decomposition, we can write the position
operator $X$ as
\begin{equation}\label{defX}
  X= \sum_{x=0}^{N-1} a(x-j)
  \varphi_{x}\langle\varphi_{x},\cdot\rangle\ ,
\end{equation}
that clearly satisfies  $X \varphi_{x}= a(x-j) \varphi_{x}$.
Momentum is formalized in the Hilbert space by means of a basis
$\{\phi_{p}\}$, unbiased to the position basis, where $p$ is an
integer number that can take any value in the set $\{ 0, 1,
\cdots , N-1 \}$. The momentum operator is given in terms of its
spectral decomposition as
\begin{equation}\label{defP}
P =\sum_{p=0}^{N-1}g(p-j)\phi_{p}\langle\phi_{p},\cdot\rangle
 \ ,
\end{equation}
where $g$ is a constant with units of momentum. The eigenvalues
of $P$ have been defined in a way to allow for movement of the
particle in both directions, anti-clockwise (positive
eigenvalues) and clockwise (negative eigenvalues) along the
circular lattice. Notice however that the state of zero momentum
is only possible when $N$ is odd. We will find in this work that
there are several important differences in the system when $N$ is
even or odd. The position and momentum bases are related by a
unitary transformation similar to the Discrete Fourier Transform
\begin{equation}\label{fix1}
\varphi_{x} =\frac{1}{\sqrt{N}}\sum_{p=0}^{N-1}
  \omega^{-(p-j)(x-j)-\alpha(x-p)}\ \phi_{p}\ ,
\end{equation}
and
\begin{equation}\label{fip1}
\phi_{p} =\frac{1}{\sqrt{N}}\sum_{x=0}^{N-1}
  \omega^{(p-j)(x-j)+\alpha(x-p)}\ \varphi_{x}\ ,
\end{equation}
with
\begin{equation}\label{fixfip}
\langle\varphi_{x},\phi_{p}\rangle =\frac{1}{\sqrt{N}}
  \omega^{(p-j)(x-j)+\alpha(x-p)}\ ,
\end{equation}
 where $\alpha$ is a parameter such that
\begin{equation}\label{epsi}
 \alpha = \left\{
  \begin{array}{rl}
  0 & \text{for $N$ odd}, \\ \frac{1}{2} & \text{for $N$ even}.
  \end{array}
\right.
\end{equation}
The constants $a$ and $g$ are not independent but are related by
\begin{equation}\label{agNeq}
 agN= 2\pi\ .
\end{equation}
This condition follows\cite{dlt} from the requirement that in the
limit $N\rightarrow\infty$ the commutation relation of position
and momentum should be $[X,P]\rightarrow i$ (we adopt units such
that $\hbar =1$). Notice that for \emph{finite} $N$, the
commutation relation $[X,P]=i$ is impossible.
\par
All these definitions are compatible with the physical
requirement that momentum is the generator of translation and
position generates increase in momentum. That is,
\begin{equation}\label{gener}
  \begin{array}{rl}
e^{-iaP}\ \varphi_{x} & = \omega^{\alpha}\ \varphi_{[x+1]} \\
e^{igX}\ \phi_{p} & = \omega^{\alpha}\ \phi_{[p+1]} \ ,
  \end{array}
\end{equation}
where  the symbol $[\cdot]$ denotes modulo $N$, that is, $[N]=0$.
Notice that an $N$-fold application of these translation
operators is equal to the identity $\mathbf{1}$ if $N$ is odd but
is equal to $-\mathbf{1}$ if $N$ is even. This is reminiscent of
a $2\pi$ rotation of a spin $1/2$ system.
\par
In equations (\ref{fix1}), (\ref{fip1}) and (\ref{fixfip}) we
could absorb the phases $\alpha x$ and $\alpha p$ in the bases
$\{ \varphi_{x}\}$ and $\{\phi_{p}\}$ by an appropriate phase
transformation (this is only relevant for even $N$ because
$\alpha\neq 0$). However this option would result in a
complication of Eqs. (\ref{gener}) where the phase
$\omega^{\alpha}$ would not appear but a sign change would appear
in the translation from site $x=N-1$ to site $x=0$ and also from
$p=N-1$ to $p=0$ loosing thereby the homogeneity of the lattice
because not all lattice sites would be equivalent. Later in this
work it will be convenient to take this option.
\section{SPREADING OF A LOCALIZED STATE AND DIFFUSION TIME}
At any instant of time, the state of the particle will be
determined by a Hilbert space element $\Psi(t)$ that in the
position representation is given by the coefficients $c_{x}(t)$
such that
\begin{equation}\label{sta}
  \Psi(t)=\sum_{x=0}^{N-1}
  c_{x}(t)\varphi_{x}\ .
\end{equation}
A given state $\Psi(0)$ at an initial time $t=0$  will evolve
according to the time evolution unitary operator given in terms
of the hamiltonian $H$ as
\begin{equation}
U_{t} = \exp(-iHt)\ .
\end{equation}
In this work we are interested in the time evolution of a state
corresponding to a particle initially localized in a lattice
site, say at $x=0$, at rest, that is, with $\langle P\rangle=0$.
Such a state is given by $\Psi(0)=\varphi_{0}$, that is,
$c_{x}(0)=\delta_{x,0}$. Let us assume a free particle with
hamiltonian $H=P^{2}/2m$. With this hamiltonian we can easily
find that the state for any time will be given by
\begin{equation}\label{cxt}
 c_{x}(T)=\frac{1}{N}\sum_{p=0}^{N-1}
  \omega^{\left(x(p-j+\alpha)-(p-j)^{2}T\right)}\ ,
\end{equation}
where we have introduced a dimensionless time parameter
$T=t/\tau$ with  a \emph{time scale} $\tau$ defined by
\begin{equation}\label{tau}
 \tau=\frac{2ma}{g}\ ,
\end{equation}
that, as we will see later, corresponds essentially to the
diffusion time. We have chosen the free particle hamiltonian,
however many of the following results do not depend on the
specific form of this hamiltonian and are also valid for any
hamiltonian invariant under the transformation $P\rightarrow -P$.
During the time evolution of a particle, initially in the site
$x=0$, the expectation value of the position and momentum will
remain zero, $\langle X\rangle=\langle P\rangle=0$ but, due to
the quantum spreading of the state, the probability distribution
of the occupation of other lattice sites will grow. We will study
some features of this quantum diffusion. The probability of
occupation of the lattice site $x$ at time $T$ is given by
\begin{equation}\label{cxt2}
  |c_{x}(T)|^{2}=\frac{1}{N^{2}}\sum_{p=0}^{N-1}\sum_{q=0}^{N-1}
  \omega^{(p-q)x-(p-q)(p+q-2j)T}\ .
\end{equation}
One of the sums can be analytically performed after a change of
the summation indices but it is not really convenient to do it.
\par
Due to the periodicity of the lattice, we expect that the
amplitudes and probabilities of Eqs.(\ref{cxt}) and (\ref{cxt2})
will be periodic in time. This is indeed the case but with
different periodicity for $N$ even or odd. That is, for the
amplitude we have
\begin{equation}\label{period1}
 c_{x}(T)=\left\{\begin{array}{ll}
    c_{x}(T+N) &\ \mbox{ for } N  \mbox{ odd } \\
     c_{x}(T+4N) &\ \mbox{ for } N  \mbox{ even } \
  \end{array}\right. \ ,
\end{equation}
and for the probability we get
\begin{equation}\label{period2}
 |c_{x}(T)|^{2}=\left\{\begin{array}{ll}
    |c_{x}(T+N)|^{2} &\ \mbox{ for } N  \mbox{ odd } \\
    |c_{x}(T+N/2)|^{2} &\ \mbox{ for } N  \mbox{ even } \
  \end{array}\right. \ .
\end{equation}
It is remarkable that the period of the amplitude is equal to the
period of the probability  for $N$ odd, but it is eight times
longer if $N$ is even. The periodicity shown in
Eqs.(\ref{period1}) and (\ref{period2}) correspond to our
particular initial state but it follows essentially from the
hamiltonian and the cyclic relation (\ref{cycl}) and therefore
this periodicity is also valid for arbitrary states and
probabilities.
\par
From the symmetry of the lattice and of the initial state we
expect that the particle will diffuse with equal probability
clockwise or anti-clockwise, that is, $ |c_{N-x}(T)|=|c_{x}(T)|$
but for the amplitude we may have a different phase on both sides
of the initial position. We will now show that the amplitude on
lattice points symmetric with respect to the initial position
$x=0$ are related by
\begin{equation}\label{cnxcx}
 c_{N-x}(T)=\omega^{-2\alpha x}c_{x}(T)\ .
\end{equation}
In order to prove this, consider
\begin{equation}\label{prv1}
 c_{N-x}(T)=\frac{1}{N}\sum_{p=0}^{N-1}
  \omega^{\left(N(p-j+\alpha)-x(p-j+\alpha)-(p-j)^{2}T\right)}\ .
\end{equation}
Using (\ref{cycl}) we eliminate $N(p-j+\alpha)$ in the exponent
and we add $Nx$. Therefore
\begin{equation}\label{prv2}
 c_{N-x}(T)=\frac{1}{N}\sum_{p=0}^{N-1}
  \omega^{\left(x(N-p+j-\alpha)-(p-j)^{2}T\right)}\ .
\end{equation}
Now we define another summation index $q=N-p$ with values in
$\{N,N-1,\cdots,1\}$.
\begin{equation}\label{prv3}
 c_{N-x}(T)=\frac{1}{N}\sum_{q=1}^{N}
  \omega^{\left(x(q+j-\alpha)-(N-q-j)^{2}T\right)}
  \ .
\end{equation}
Since $N=2j+1$, the squared parenthesis in the exponent becomes
$(q-1-j)^{2}$. Then
\begin{equation}\label{prv4}
 c_{N-x}(T)=\frac{\omega^{x(1+2j-2\alpha)}}{N}\sum_{q=1}^{N}
  \omega^{\left(x(q-1-j+\alpha)-(q-1-j)^{2}T\right)}\ .
\end{equation}
Finally, redefining the summation index $p=q-1$ and using again
(\ref{cycl}), we find that the right hand side of this equation
is $\omega^{-2\alpha x}c_{x}(T)$. A remarkable consequence of
relation (\ref{cnxcx}) is that, \emph{for even $N$, a quantum
particle in a localized state will never diffuse to the antipode
location.} The antipode location, $x=N/2$, exists only for even
$N$. The proof follows from Eq.(\ref{cnxcx}) since we have
$c_{N-N/2}=\omega^{-N/2}c_{N/2}$, but $\omega^{-N/2}=-1$
therefore $c_{N/2}=-c_{N/2}$. That is,
\begin{equation}\label{antipo}
 c_{N/2}(T)=0\ \forall\, T\ .
\end{equation}
This is a remarkable result that can be checked by explicit
evaluation from Eq.(\ref{cxt}) redefining the summation index
$q=p-j$ running from $-j$ to $j$. Doing this we obtain a sum
whose terms are anti-symmetric under $q\rightarrow-q$; therefore
they add to zero. Another physically appealing proof of this
result is provided by Feynman's ``sum over paths''
method.\cite{fey} In this case, a path contributing to the
probability amplitude for the transition from $x=0$ at $t_{i}$ to
$x=N/2$ at $t_{f}$ is defined by a set $\{x_{k};t_{k}\}$ for each
partition of the time interval $t_{i}<t_{k}<t_{f}$. It turns out
that for each path $\{x_{k};t_{k}\}$ going from $x=0$ to $x=N/2$
there is another path $\{z_{k};t_{k}\}$, symmetric with respect
to $x=0$, that is, $z_{k}=N-x_{k}$ but with the same values of
$\{t_{k}\}$, that cancels its contribution to the probability
amplitude, simply because $dz_{k}=-dx_{k}$. In the case $N=2$,
besides the initial location, there is only one remaining
location, the antipode. Therefore, for all time, the particle
will remain in its initial position. Clearly, for $N=2$ the
states $\varphi_{0}$ and $\varphi_{1}$ are not only position
eigenvectors but  also eigenstates of the hamiltonian and
therefore they are stationary states. For odd values of $N$ the
antipode does not exists but we can study the transition
probability to diffuse to the ``farthest'' locations $x=(N \pm
1)/2$. We will later see a remarkable difference in the odd-$N$
case. We will see that, contrary to what happens in the even case
in which the antipodes are never reached, if $N$ is odd a sharp
distribution will build up in an environment of the antipode at
the time $T=N/2$. This is precisely the time when the state is
reconstructed in the $N$-even case but at the original site.
\par
We will now calculate the \emph{diffusion time}, that is, the
time that is required for a particle, initially localized  in one
lattice site, to ``diffuse'' to the whole cyclic lattice. Since
the state is periodic in time, with period proportional to $N$,
we expect that the state reconstruction happens after the whole
lattice is visited and therefore the diffusion time should be, at
most, proportional to $N$. In order to calculate the diffusion
time we must find the time dependence of the  width of the
probability distribution of position. It turns out that for
finite $N$, or for periodic distributions, the quantities
$\langle\Psi(t), X\Psi(t)\rangle$ and $\langle \Psi(t),
X^{2}\Psi(t)\rangle-\langle\Psi(t), X\Psi(t)\rangle^{2}$
\emph{are not} appropriate estimates for the center
$\overline{X}$ and width $\Delta$ of the distribution along a
cyclic lattice or ring. The main reason why they are not
appropriate is that any physical quantity in a cyclic lattice
should be periodic, that is, invariant under $x\rightarrow x+Na$
and clearly the quantity $\langle\Psi(t), X\Psi(t)\rangle$ does
not complies to this. Two simple examples: first, let us suppose
a distribution given by
$|c_{N-1}|^{2}=|c_{0}|^{2}=|c_{1}|^{2}=1/3$. Clearly the center
of the distribution is at the location corresponding to the label
$x=0$, that is, at the position $-aj=-a(N-1)/2$, but the quantity
$\langle\Psi(t), X\Psi(t)\rangle$ is
$\sum_{x=0}^{N-1}a(x-j)|c_{x}|^{2}=-a(N-3)/6$. For another
example, consider a uniform distribution that fills completely a
ring. In our case of a cyclic lattice we have $|c_{x}|^{2}=1/N,\
\forall x$. Clearly, this distribution \emph{does not have a
center;} it should be undefined on the ring because all points
are equivalent, but the quantity $\langle\Psi(t),
X\Psi(t)\rangle$ is $\sum_{x=0}^{N-1}a(x-j)|c_{x}|^{2}=0 $.
\par
The problem of defining the center $\overline{X}$ and width
$\Delta$ of a distribution in a ring or cyclic lattice has been
solved\cite{for1,for2} using the concept of the \emph{ centroid}
of a distribution on a ring. Let us build a map of the ring into
a unit circle in the complex plane. In order to define the
centroid $Z$ for a probability distribution $|c_{x}|^{2}$ on the
sites $x=0,1,\cdots,N-1$ of a cyclic lattice, let us consider the
unit circle in the complex plane with $N$ points located at
$\omega^{x}$. The centroid of the distribution is a complex
number $Z=\rho e^{i\theta}$ given by
\begin{equation}\label{centroid}
  Z=\rho e^{i\theta}=\sum_{x=0}^{N-1}\omega^{x}|c_{x}|^{2}\ .
\end{equation}
The radial projection of the centroid on the unit circle maps the
center of the distribution on the ring. Therefore,
\begin{equation}\label{xcent}
   \overline{X}=a(\frac{\theta}{2\pi}N-j)\ ,
\end{equation}
and the width $\Delta$ of the distribution is given by
\begin{equation}\label{width}
 \Delta^{2}=(aN)^{2}(1-|Z|^{2})\ ,
\end{equation}
where the factor $aN$ has been chosen such that for a uniform
distribution covering the whole lattice ($Z=0$) the width of the
distribution is equal to the size of the lattice. Notice that
when only one site is occupied, the width is zero. These
definitions are clarified in an example shown in Figure 2.
\par
We can now study the time dependence of the width for our initial
condition of a particle at rest, localized in $x=0$. Let us
calculate first the centroid. From Eqs.(\ref{centroid}) and
(\ref{cxt2}) we obtain
\begin{equation}\label{centrT}
Z=\frac{1}{N^{2}}\sum_{p=0}^{N-1}\sum_{q=0}^{N-1}\sum_{x=0}^{N-1}\omega^{x}
  \omega^{(p-q)x-(p-q)(p+q-2j)T}\ .
\end{equation}
The sum over $x$ can be performed:
\begin{equation}\label{sumx}
  \sum_{x=0}^{N-1}\omega^{(p-q+1)x}=N(\delta_{q,p+1}+
  \delta_{q,0}\delta_{p,N-1})\ .
\end{equation}
The first term of the parenthesis corresponds to the vanishing of
the exponent of $\omega$ and the second term is for the case when
the exponent is equal to $Nx$. With the Kronecker deltas we
perform the sum over $q$, and the remaining sum over $p$ has a
known result. We get then,
\begin{equation}\label{centrT1}
Z=\frac{1}{N}\left(\frac{\sin\left(\frac{2\pi}{N}(N-1)T\right)}
{\sin\left(\frac{2\pi}{N}T\right)}+1\right)\ .
\end{equation}
As expected, the centroid has the same time periodicity as the
probability distribution, that is, $N$ for odd number of lattice
sites and $N/2$ for an even number of sites. Due to the initial
condition of a particle in the site $x=0$ and to the symmetric
diffusion, the centroid is real at all times. The study of the
time dependence of the centroid, shown in Figs. 3a and 3b for
$N=16$ and $17$, allows a simple qualitative description of the
time evolution of the distribution. In the figures we notice that
the centroid oscillates most of the time with values close to
zero, corresponding to distributions close to (but not
necessarily equal to) uniform distributions covering the whole
lattice. At time $T=N/2$ the centroid assumes the value of $Z=1$
in the $N$-even case, as expected because at this time the
initial state is reconstructed, and for odd $N$ it takes the
value $-(N-2)/N$, close to $Z=-1$ for large $N$, implying that at
the time $T=N/2$ the distribution is concentrated at the
antipodes of the initial location; however, precisely at the
antipode there is no lattice site for odd $N$ and the state can
not be reconstructed in one location. We see here a sharp
distinction in the behaviour of diffusion in the even and odd
case: the antipode is never reached in the $N$-even case but the
distribution peaks in the neighbourhood of the antipode (at time
$T=N/2$) in the $N$-odd case.
\par
With the knowledge of the centroid, we can now calculate the time
dependence of the width of the distribution. In particular we
want to find the \emph{diffusion time} $T_{D}$, that we define as
the time when the width assumes its maximal value $aN$ for the
first time. Notice that when the centroid vanishes, the width
assumes its maximal value. From Eq.(\ref{centrT1}) we see that
the centroid vanishes for $T=1$ for all $N$, therefore the width
is maximal ($aN$) at $T=1$. However Eq.(\ref{centrT1}) has
another root for a time $T$ smaller than $1$ when $N>4$. Of
course, when $N=2$ the diffusion time is infinite because the
particle never diffuses out of the initial site. Summarizing we
have
\begin{equation}\label{covT}
T_{D}=\left\{\begin{array}{ll}
    \infty &\ , \mbox{ for } N=2   \\
     1 &\ ,\mbox{ for } N=3 \\
     \frac{N}{2(N-2)}&\ ,\mbox{ for } N\geq 4  \
  \end{array}\right. \ .
\end{equation}
It might at first seem strange that the defined diffusion time
\emph{decreases} towards a constant value $T_{D}=1/2$ with an
\emph{increasing} number of sites $N$, but we can see that this
is to be expected as a consequence of indeterminacy principle.
Increasing the number of sites $N$ with the same initial
condition of a particle in one site, is equivalent to a sharper
localization of the initial state. This implies a wider momentum
spread, responsible for a faster diffusion that decreases the
diffusion time. The explicit time dependence of the width of the
distribution is then given by
\begin{equation}\label{widthT}
  \Delta=a\sqrt{N^{2}-\left(\frac{\sin\left(\frac{2\pi}{N}(N-1)T\right)}
{\sin\left(\frac{2\pi}{N}T\right)}+1\right)^{2}}\ .
\end{equation}
This quantity is zero at $T=0$, grows with time, and takes the
maximal value $aN$ at $T=T_{D}$; then it oscillates with values
close to the maximal value except at time $T=N/2$ when the width
becomes zero for $N$-even or decreases to $a2\sqrt{N-1}$ for odd
$N$. At this time, that we call \emph{first reconstruction time}
$T_{R}=N/2$, the particle is reconstructed at the original site
($N$ even) or is concentrated near the antipode ($N$ odd). Notice
that at this reconstruction time $T_{R}$ the \emph{state} is
reconstructed only if $N$ is even whereas for odd $N$ the
probability distribution for the location of the particle peaks
but there is no exact reconstruction of the particle in one
location of the antipode. For very short times $T\ll T_{D}$ the
system does not notices the geometry of the cyclic lattice and
the width grows linearly with time with a diffusion speed
\emph{increasing} with the lattice size $N$. Indeed, the first
term in the Taylor expansion of $\Delta$ is
\begin{equation}\label{widthTshort}
 \Delta=a2\pi\sqrt{\case{1}{3}(N-1)(N-2)}\ T\ \mbox{ for } T\ll T_{D} \ .
\end{equation}
\par
We can now investigate whether the reconstruction of a localized
state for the particle at time $T_{R}=N/2$ at the original site
($N$ even) or the concentration of the particle near the
antipodes ($N$ odd) is affected by the parity of the initial
state. The initial state considered above, a particle in
\emph{one site}, has necessarily even parity. In order to be able
to study also the effect of an odd parity initial state we will
consider an initial state of a particle at rest, $\langle
P\rangle =0$, in an even or odd superposition of two neighbouring
position eigenstates corresponding to the sites $x=0$ and $x=1$:
\begin{equation}\label{psievenodd}
 \Psi_{{\pm}}(0)=\case{1}{\sqrt{2}}\ \left(\varphi_{0}\pm\omega^{\alpha}\varphi_{1}\right)
\end{equation}
With this initial state, we can calculate the time evolution of
the centroid. However, the centroid will no longer be a real
number. It is therefore convenient to make a rotation of the
centroid in the complex plane by an angle $\omega^{-1/2}$ in
order to obtain the real quantity $\widetilde{Z}_{\pm}(T) =
\omega^{-1/2}Z_{\pm}(T)$, where $Z_{\pm}(T)$ is the centroid
corresponding to the two initial states $\Psi_{\pm}(0)$. This results
in
\begin{eqnarray}\label{Pseudocentr}
\nonumber
  \widetilde{Z}_{\pm}(T)&=&  \frac{1}{N}\left(
\cos\left(\frac{\pi}{N}\right)\frac{\sin\left(\frac{\pi}{N}(N-1)2T\right)}
{\sin\left(\frac{\pi}{N}2T\right)}
  \pm\frac{\sin\left(\frac{\pi}{N}(N-1)(2T-1)\right)}
{2\sin\left(\frac{\pi}{N}(2T-1)\right)}\right. \\
   &\pm & \left.\frac{\sin\left(\frac{\pi}{N}(N-1)(2T+1)\right)}
{2\sin\left(\frac{\pi}{N}(2T+1)\right)}
 +\cos\left(\frac{\pi}{N}\right)\mp 1\right)\ .
\end{eqnarray}
In Figures 4a and 4b we see the time evolution of the (rotated)
centroid $ \widetilde{Z}_+(T)$ for an even initial state
$\Psi_{+}(0)$ for even and odd $N$. For a qualitative comparison
with Figure 1, we have taken $N=33$ and $N=34$ in order to have
similar relation between the size of the lattice and the number
of sites of the initial state. From this comparison it is clear
that the behaviour is similar. At time $T=N/2$, a localized
\emph{even} state is reconstructed at the original locations if
$N$ is even  or the particle is localized at the antipodes if $N$
is odd. In Figures 5a and 5b we can see that this is also true
when the initial state $\Psi_{-}(0)$ is \emph{odd}, but the
effect is much blurred by rapid oscillations of the centroid.
Shortly before and after every reconstruction of the particle, it
is \emph{almost} reconstructed but on the opposite side of the
lattice. For both, even and odd parity states, the initial value
of the centroid, $\widetilde{Z}_{\pm}(0)=\cos(\pi/N)$, is exactly
recovered for even $N$ at time $T=N/2$ (this must be so because
the state is periodic) and for odd $N$ the centroid reaches the
minimum value $\widetilde{Z}_{\pm}(N/2)=-((N-2) \cos(\pi/N) \pm
2)/N$. For large $N$ this minimum value approaches
$-\cos(\pi/N)$, corresponding to the occupation of two
neighbouring sites at the antipodes.
\section{THE CONTINUOUS LIMIT}
We have found that there are very strong differences  in the
behaviour of the system when $N$ takes even or odd values. Of
course, all these differences must be compatible with the
continuous limit when $N\rightarrow\infty$ where we can not
differentiate between even or odd $N$. In this section we will
investigate this limit. First we must redefine the indices of
summation in a symmetric way such that they can take positive and
negative values. Let then
\begin{equation}\label{simetrvar}
  \begin{array}{ll}
   y=a(x-j) & \in [-aj,aj] \\
   q=g(p-j)&  \in [-gj,gj]\ .
  \end{array}
\end{equation}
Anticipating that in the limit $N\rightarrow\infty$ the position
and momentum eigenfunctions will not be normalizable, we define
these eigenfunctions in terms of the symmetric indices as
\begin{equation}\label{eigenf}
  \varphi_{y} =\frac{1}{\sqrt{a}}\ \varphi_{x}\ \mbox{ and }\
\phi_{q} =\frac{1}{\sqrt{g}}\ \phi_{p} \ .
\end{equation}
If in the limit $N\rightarrow\infty$ we also take $a\rightarrow
0$ or $g\rightarrow 0$ then the summations become integrals
according to the scheme
\begin{equation}\label{sumint}
  \sum_{y=-aj}^{aj}\!\!\! a \rightarrow \int^{\infty}_{-\infty}\!\!\! dy \ \mbox{ or }
 \sum_{q=-gj}^{gj}\!\!\! g \rightarrow \int^{\infty}_{-\infty}\!\!\! dq \ .
\end{equation}

The limit $N\rightarrow\infty$ is constrained by the condition
$Nag=2\pi$ and therefore we will consider three different limits
$L1,\ L2,\ L3,$ that will correspond to three different physical
systems: \
\begin{equation}\label{L1}
  L1:\left\{
  \begin{array}{l}
   N\rightarrow\infty,\ a\rightarrow 0,\ g\rightarrow 0,\\
    y\in [-\infty,\infty],\  q\in [-\infty,\infty] \ .
    \end{array}
    \right.
\end{equation}
\begin{equation}\label{L2}
   L2:\left\{
\begin{array}{l}
    N\rightarrow\infty,\ a\rightarrow 0,\ Na=L,\
   g=\frac{2\pi}{L},\\
    y\in [-L/2,L/2],\  q= \frac{2\pi}{L} n,\  n=\left\{
  \begin{array}{ll}
  \pm 1/2, \pm 3/2, \cdots & N \text{even} ,  \\
  0,\pm 1, \pm 2, \cdots  & N \text{odd}.
  \end{array} \right.\end{array}\right.
  \end{equation}
   \begin{equation}\label{L3}
   L3: \left\{
\begin{array}{l}
 N\rightarrow\infty,\ g\rightarrow 0,\ Ng=G,\
   a=\frac{2\pi}{G},\\
   q\in [-G/2,G/2],\ y= \frac{2\pi}{G} n,\ n=\left\{
  \begin{array}{ll}
  \pm 1/2, \pm 3/2, \cdots & N \text{even} ,  \\
  0,\pm 1, \pm 2, \cdots  &N \text{odd}.
 \end{array}
 \right.\end{array}\right.
\end{equation}
In the limit $L1$ both variables $y$ and $q$ are continuous and
unbound whereas in $L2$ the variable $y$ is bounded and
continuous but $q$ is unbound and discrete; these properties are
exchanged in $L3$.

In the limit $L1$, the physical system becomes a free particle
moving in a one dimensional infinite space where position and
momentum observable can take continuous values. In the limit
$L2$, the physical system is a free particle moving in a ring of
perimeter $L$. Position is continuous and takes values from
$-L/2$ to $L/2$ whereas momentum is a discrete variable.  We will
later see that among the two choices for the number $n$, only the
values $0,\pm 1, \pm 2, \cdots$ are physically meaningful. This
system also corresponds to a particle in a box with periodic
boundary conditions.  Finally, in the limit $L3$, the physical
system is a particle moving in a one dimensional infinite lattice
with lattice constant $a=2\pi/G$ and continuous momentum
restricted to the Brillouin zone $[-G/2,G/2]$.

The striking differences in the behaviour of the system between
even and odd $N$ appear in the time periodicity of the state and
probability, and in the first reconstruction time for the
probability distribution. These differences involve a time scale
$t=N\tau\propto Na/g =2\pi/g^{2}$. In both limits, $L1$ and $L3$,
this time scale is \emph{infinite} and therefore we should not
worry about whether $N$ is even or odd when taking the limit
$N\rightarrow\infty$; however in the limit $L2$ the time scale is
\emph{finite} and proportional to $L^{2}$. In this last case we
will see that the even $N$ case is mathematically sound but does
not corresponds to any reasonable physical system.

It is convenient, in order to analyse the $L1$ and $L2$ limits,
to adopt the position representation of the eigenfunctions where
the momentum eigenvectors are given by
Eqs.(\ref{fixfip},\ref{eigenf}) as
\begin{equation}\label{posrep}
    \phi_{q}(y) =\langle\varphi_{y},\phi_{q}\rangle =
\frac{1}{\sqrt{2\pi}}\ e^{iqy+i\alpha(gy-aq)}\ .
\end{equation}
In the limit $L1$, where $a\rightarrow 0$ and $g\rightarrow 0$,
this eigenfunction becomes
\begin{equation}\label{posrep1}
    \phi_{q}(y) =
\frac{1}{\sqrt{2\pi}}\ e^{iqy}\ ,
\end{equation}
provided that, in the even $N$  case ($\alpha=1/2$), The values
of $y$ and $q$ remain finite (otherwise a minus sign can appear).
Expanding the position eigenfunctions in the momentum basis we
obtain
\begin{equation}\label{posrep2}
    \varphi_{y'}(y) =
\frac{1}{2\pi}\int_{-\infty}^{\infty}\!\!\!dq\ \
e^{iq(y-y')}=\delta(y-y')\ .
\end{equation}
We obtain therefore the usual position and momentum
eigenfunctions for a free particle moving in a line.

Let us now consider the $L2$ limit where we have two
possibilities: $\alpha=0,\ n=0,\pm 1, \pm 2,\ \cdots $
 and $\alpha=1/2,\ n=\pm 1/2,\ \pm 3/2,\ \cdots $. In the first
case Eq.(\ref{posrep}) results in
\begin{equation}\label{posrep3}
  \phi_{q}(y) =
\frac{1}{\sqrt{2\pi}}\ e^{iy\frac{2\pi}{L}n}\ ,\ n=0,\pm 1, \pm
2,\ \cdots\ ,
\end{equation}
and in the second case, assuming $m=\pm 1/2,\ \pm 3/2,\ \cdots $,
we have
\begin{equation}\label{posrep4}
  \phi_{q}(y) =
\frac{1}{\sqrt{2\pi}}\
e^{iy\frac{2\pi}{L}m+i\frac{1}{2}\frac{2\pi}{L}y} =
\frac{1}{\sqrt{2\pi}}\ e^{iy\frac{2\pi}{L}(m+\frac{1}{2})}=
\frac{1}{\sqrt{2\pi}}\ e^{iy\frac{2\pi}{L}n} \ ,\ n=0,\pm 1, \pm
2,\ \cdots\ .
\end{equation}
Therefore \emph{both} cases lead to the same position
representation of the momentum eigenfunction. So far it would
seem that the even and odd $N$ cases are identical in the limit
$N\rightarrow\infty$ however this is not so as we will see next.
It turns out that in the even $N$ case, when $\alpha=1/2$, the
momentum operator in the position representation \emph{is not}
given by the derivative operator as usual. In order to prove
this, consider the first equation in (\ref{gener}) written in
terms of the symmetric variables, that is,
\begin{equation}\label{gener1}
 e^{-iaP}\ \varphi_{y}  = e^{i\frac{2\pi}{N}\alpha}\
\varphi_{[y+a]}=e^{i\frac{2\pi}{Na}a\alpha}\ \varphi_{[y+a]}\ .
\end{equation}
Applying the limit $L2$ we get,
\begin{equation}\label{gener2}
 (1-iaP)\ \varphi_{y} = (1+i\frac{2\pi}{L}a\alpha)\ \varphi_{y+a}\ .
\end{equation}
Therefore
\begin{equation}\label{gener3}
 P\ \varphi_{y} = i\lim_{a\rightarrow 0}
\frac{\varphi_{y+a}-\varphi_{y}}{a}-\frac{2\pi}{L}\alpha\
\varphi_{y+a}\ ,
\end{equation}
where we see that only in the odd $N$ case, when $\alpha=0$, is
the momentum operator given by the derivative operator.

The inadequacy of even $N$ in the limit is more conveniently seen
if we absorb the phase $e^{i\alpha(gy-aq)}$ in the eigenfunctions
as was mentioned at the end of section II. In this case the
$\alpha$-dependent phase in Eq.(\ref{posrep}) would not appear
and we would have two different position representations of the
momentum eigenfunctions given by
\begin{eqnarray}
\nonumber
  \phi_{q}^{1}(y) &=& \frac{1}{\sqrt{2\pi}}\
e^{iy\frac{2\pi}{L}(0,\ \pm 1,\ \pm 2,\ \cdots)} \mbox{ for odd }N \\
  \phi_{q}^{2}(y) &=& \frac{1}{\sqrt{2\pi}}\
e^{iy\frac{2\pi}{L}(\pm 1/2,\ \pm 3/2,\ \cdots)} \mbox{ for even
}N\ .\nonumber
\end{eqnarray}
The momentum eigenfunctions $\phi_{q}^{1}(y)$ are the same as the
ones obtained before in Eq.(\ref{posrep3}) and the other ones,
$\phi_{q}^{2}(y)$, are mathematically sound but are inadequate
for physical systems because they are anti-symmetric,
$\phi_{q}^{2}(-L/2)=-\phi_{q}^{2}(L/2)$ and have period $2L$,
whereas all reasonable physical states for a particle in a ring
are symmetric and have space periodicity $L$.

As a further confirmation that the $L2$ limit corresponds with
the odd $N$ case, we will show that an initial state in a ring is
reconstructed at the antipodes at the reconstruction time
$t_{R}=T_{R}\tau=N\tau/2=mL^{2}/(2\pi)$ as it happens in the case
of finite but odd $N$. In order to prove this we assume an
arbitrary initial state expanded in terms of the momentum base
\begin{equation}\label{ringosc}
    \psi(y,0)=\sum_{q}c_{q}\ \phi_{q}(y)\ .
\end{equation}
We apply the time evolution operator to this state, considering
that
\begin{equation}\label{rinosc1}
    e^{-i\frac{P^{2}}{2m}t}\ \phi_{q}(y)=
e^{-i\frac{q^{2}}{2m}t}\ \phi_{q}(y)\ ,
\end{equation}
and using Eq.(\ref{posrep3}) we get
\begin{equation}\label{ringosc2}
    \psi(y,t)=\frac{1}{\sqrt{2\pi}}\sum_{n=0,\pm 1,\pm2,
\cdots}c_{n}\
e^{-\frac{i}{2m}\left(\frac{2\pi}{L}\right)^{2}n^{2}t
+iy\frac{2\pi}{L}n} \ .
\end{equation}
Consider now this state at the reconstruction time
$t_{R}=mL^{2}/(2\pi)$,
\begin{equation}\label{ringosc3}
    \psi(y,t_{R})=\frac{1}{\sqrt{2\pi}}\sum_{n=0,\pm 1,\pm2,
\cdots}c_{n}\ e^{-i\pi n^{2}}e^{iy\frac{2\pi}{L}n} \ .
\end{equation}
Now, since $n^{2}$ and $n$ have the same parity, it is $e^{-i\pi
n^{2}}=e^{-i\pi n}$, and we get,
\begin{equation}\label{ringosc4}
    \psi(y,t_{R})=\frac{1}{\sqrt{2\pi}}\sum_{n=0,\pm 1,\pm2,
\cdots}c_{n}\ e^{i(y-\frac{L}{2})\frac{2\pi}{L}n} \ .
\end{equation}
Therefore
\begin{equation}\label{ringosc5}
    \psi(y,t_{R})=\psi(y-\frac{L}{2},0)\ .
\end{equation}
with the meaning that the state at the reconstruction time
$t_{R}$ is equal to the initial state, but shifted to the
antipode $y-L/2$.

A particle in a localized state in a continuous ring will flip
back and forth between the original position and its antipode. If
the particle is electrically charged, the system will radiate
electromagnetic energy and decay to a nonlocalized stationary
state (not necessarily the ground state) and a small fluctuation
away of a nonlocalized state will grow, absorbing electromagnetic
energy of the appropriate frequency (for an electron in a ring of
$10\mu$ to $100\mu$ of perimeter, the radiation will be in the
radio frequencies). This radiation, or the corresponding
absorption, could be experimentally detected, in particular if
one builds a material with a large number of conducting rings.
Such a material, whose dielectric properties follow from a
fundamental quantum mechanical effect, could find technological
applications.

Finally, the $L3$ limit is treated equal to the $L2$ case but in
terms of the momentum representation of the eigenfunctions.
Similar arguments show that the even $N$ case leads, in the
limit, to unphysical situations.
\section{CONCLUSION}
In this work we have studied the diffusion of a quantum
mechanical particle, initially localized, in a ring with $N$
sites. This diffusion has qualitative features quite different
from the diffusion of a particle performing a classical random
walk. It is well known that in a classical random walk, the width
of the distribution grows like $\sqrt{T}$ whereas quantum
mechanical diffusion grows initially proportional to $T$.
Furthermore, we see in Eq. (\ref{widthTshort}) that the speed of
quantum diffusion, for large $N$, increases linearly with the
size of the lattice $aN$. This non local effect is contrary to
the classical behaviour and can be understood qualitatively as a
consequence of Heisenberg's indeterminacy principle: if the
initial state is a particle in \emph{one site} of the lattice,
increasing the number of sites is equivalent to a sharper
localization relative to the lattice size, and this results in a
wider momentum spread, responsible for the increase in diffusion
speed. Since the diffusion speed increases with the number of
sites $N$ it is reasonable to expect that the time necessary to
diffuse to the whole lattice will be constant independent of the
lattice size. This is indeed the result shown in Eq.(\ref{covT})
where we see that, for large $N$, the diffusion time $T_{D}$ is
constant. This is again in contradiction with the behaviour of
the classical random walk where the covering time\cite{tcov} (the
time it takes for a random walk to visit all the lattice sites)
for a cyclic lattice increases quadratically with $N$ (precisely,
$N(N-1)/2$).
\begin{acknowledgements}
This work received partial support from ``Consejo Nacional de
Investigaciones Cient\'{\i}ficas y T\'ecnicas'' (CONICET) and
from ADPCyT (Picto 03-08431), Argentina.
\end{acknowledgements}

\newpage \noindent
\textbf{FIGURE CAPTIONS}\\

\noindent FIGURE 1. Cyclic lattice with $N$ sites characterized
by a label $x$ running from $x=0$ to $x=N-1$ and lattice constant
$a$. The position observable corresponding to site $x$ has the
eigenvalue $a(x-j)$ where $j=(N-1)/2$ and can take positive and
negative values.\\

\noindent FIGURE 2. An example for the centroid of a
distribution. The symbol $\oplus$ shows the position of the
centroid $Z=\rho e^{i\theta}$ for a distribution where the filled
dots have a constant occupation probability and all other sites
are empty. The center $\overline{X}$ of the distribution is shown
and the
width $\Delta$ is proportional to the chord $C$ shown.\\

\noindent FIGURE 3. Time dependence of the centroid for even (a)
and odd (b) number of sites. At time $N/2$ the state is
reconstructed at the original site for even $N$ and is
concentrated at the antipodes for odd $N$.\\

\noindent FIGURE 4. Time dependence of the (rotated) centroid for
even (a) and odd (b) number of sites for an initial even state
occupying two neighbouring sites. At time $N/2$ the state is
reconstructed at the original sites for even $N$ and is
concentrated at the
antipodes for odd $N$.\\

\noindent FIGURE 5. Time dependence of the (rotated) centroid for
even (a) and odd (b) number of sites for an initial odd state
occupying two neighbouring sites. At time $N/2$ the state is
reconstructed at the original sites for even $N$ and is
concentrated at the antipodes for odd $N$.
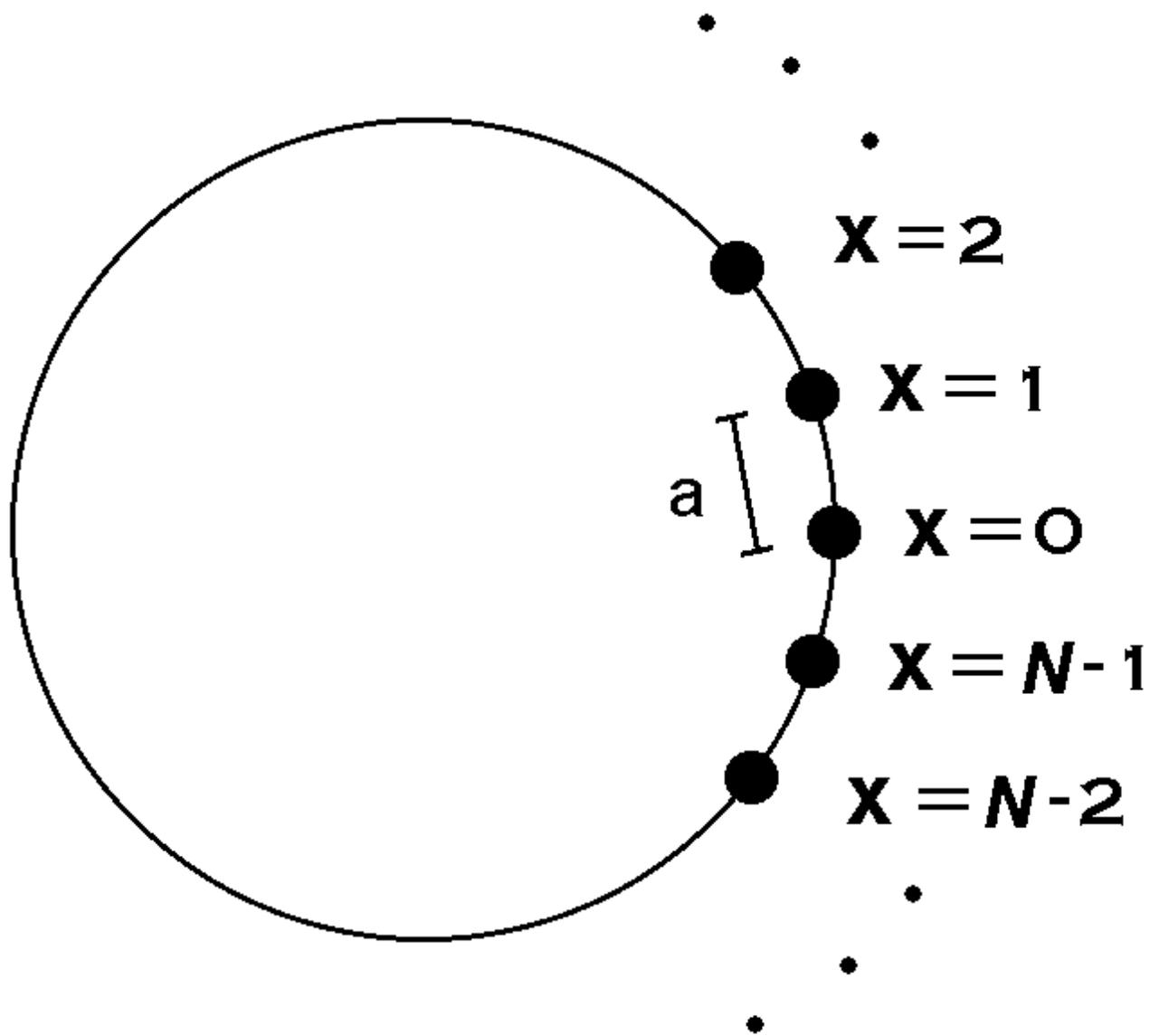
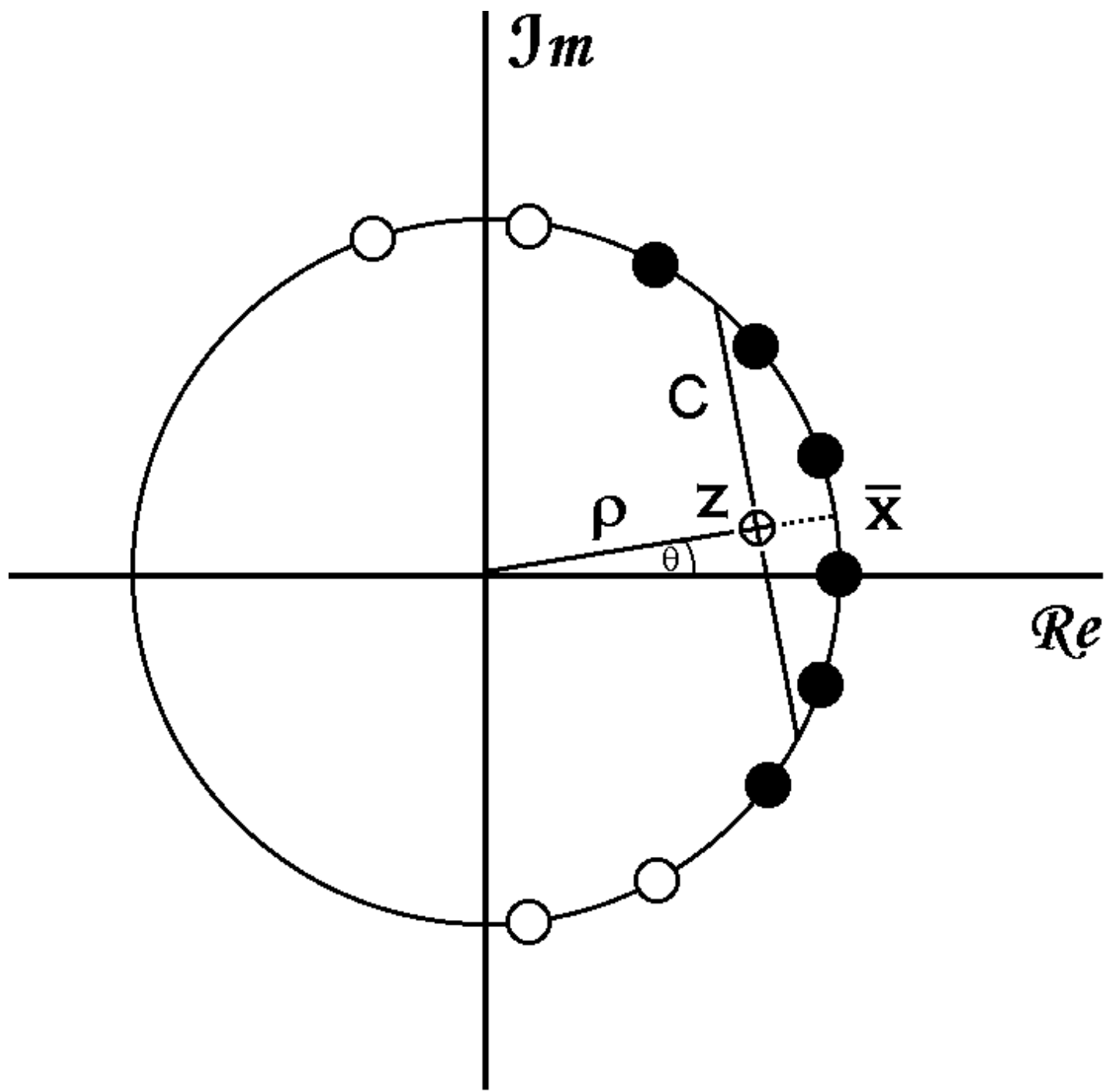
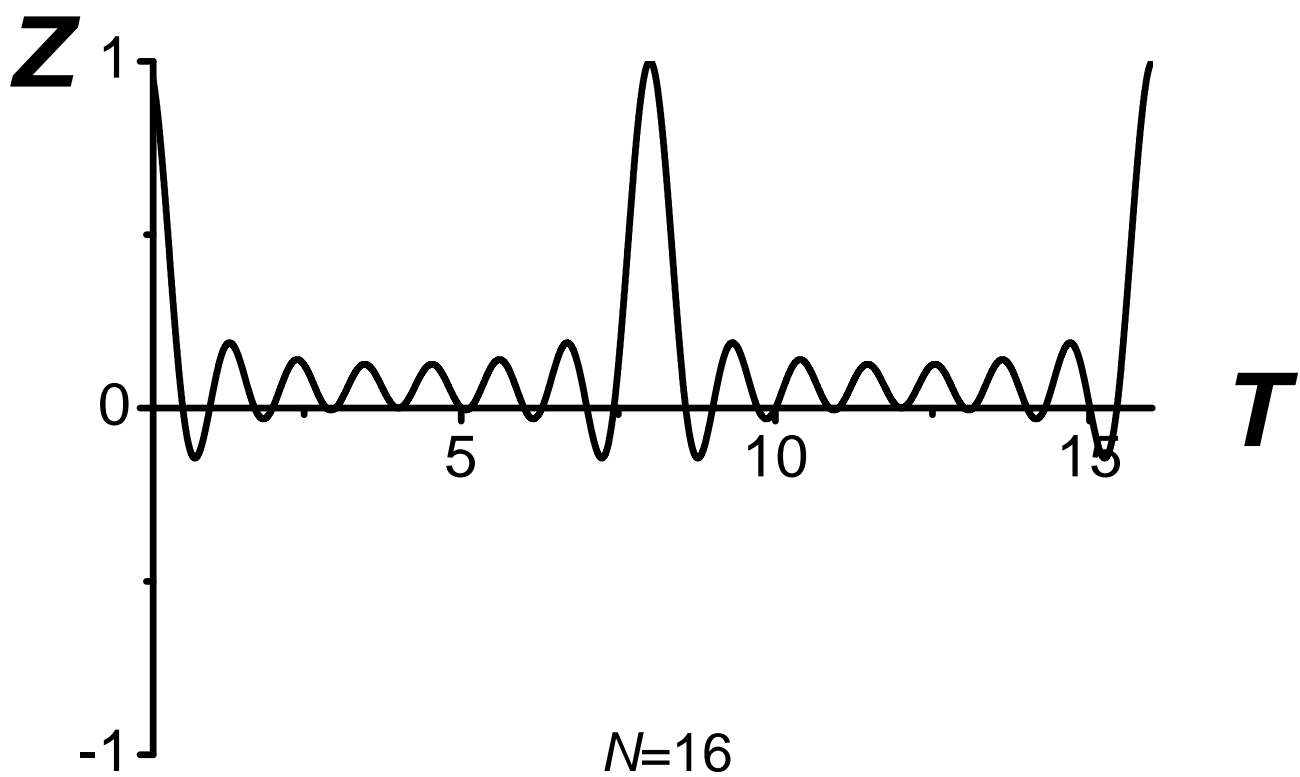
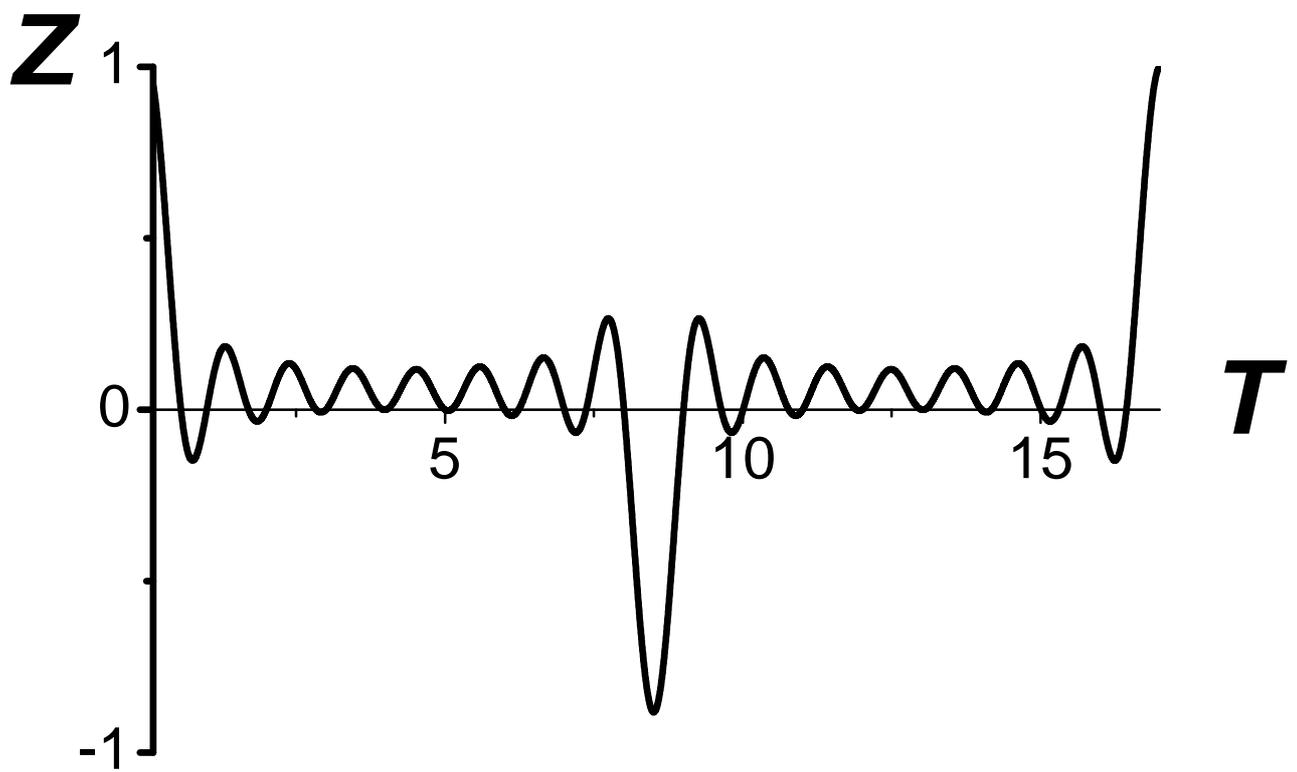
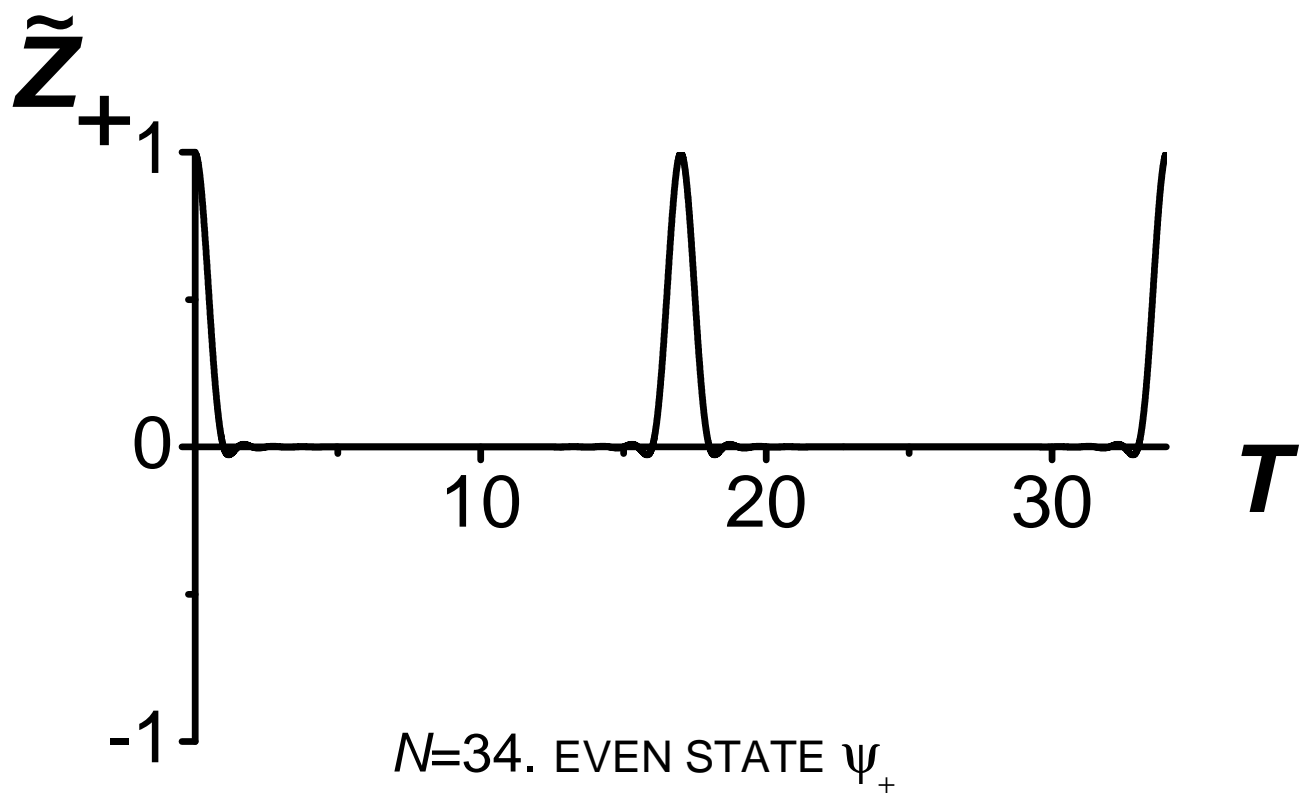
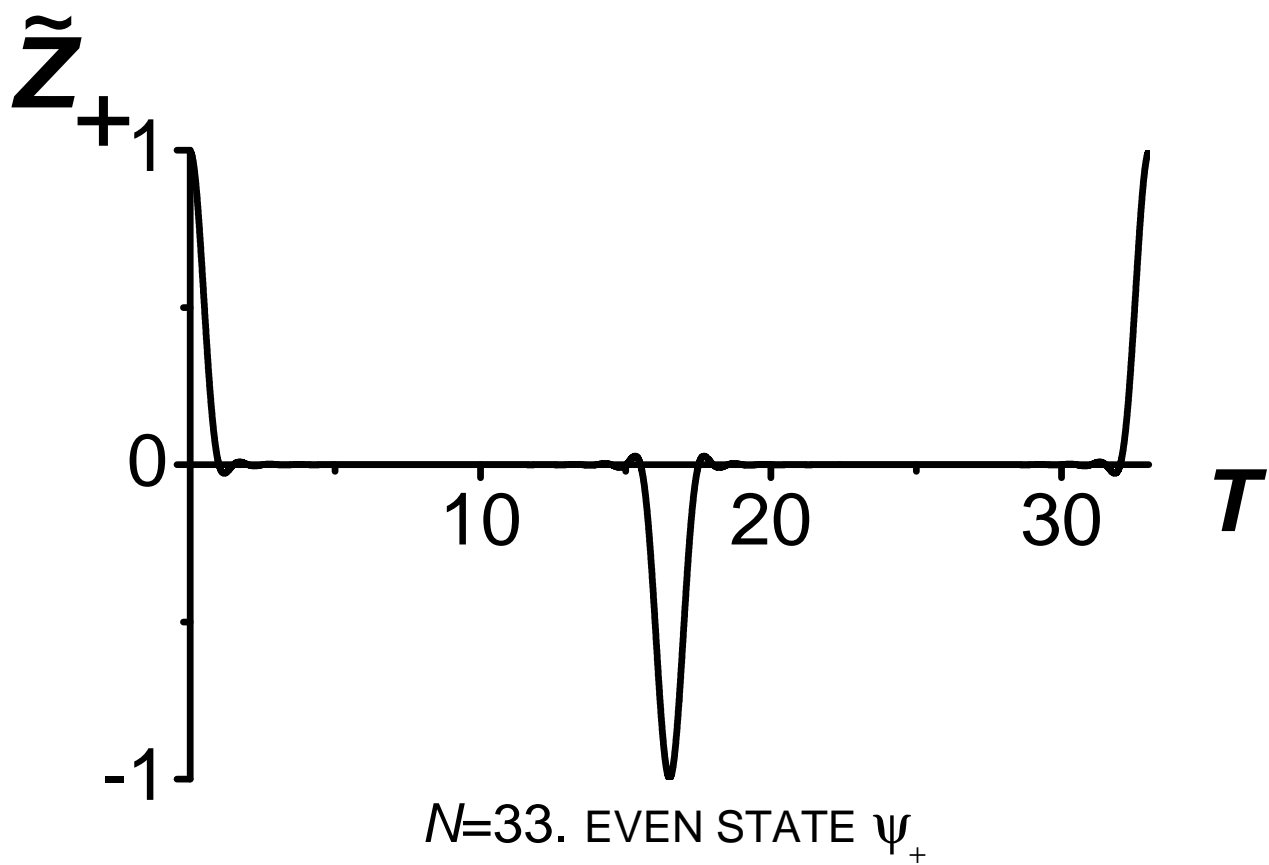
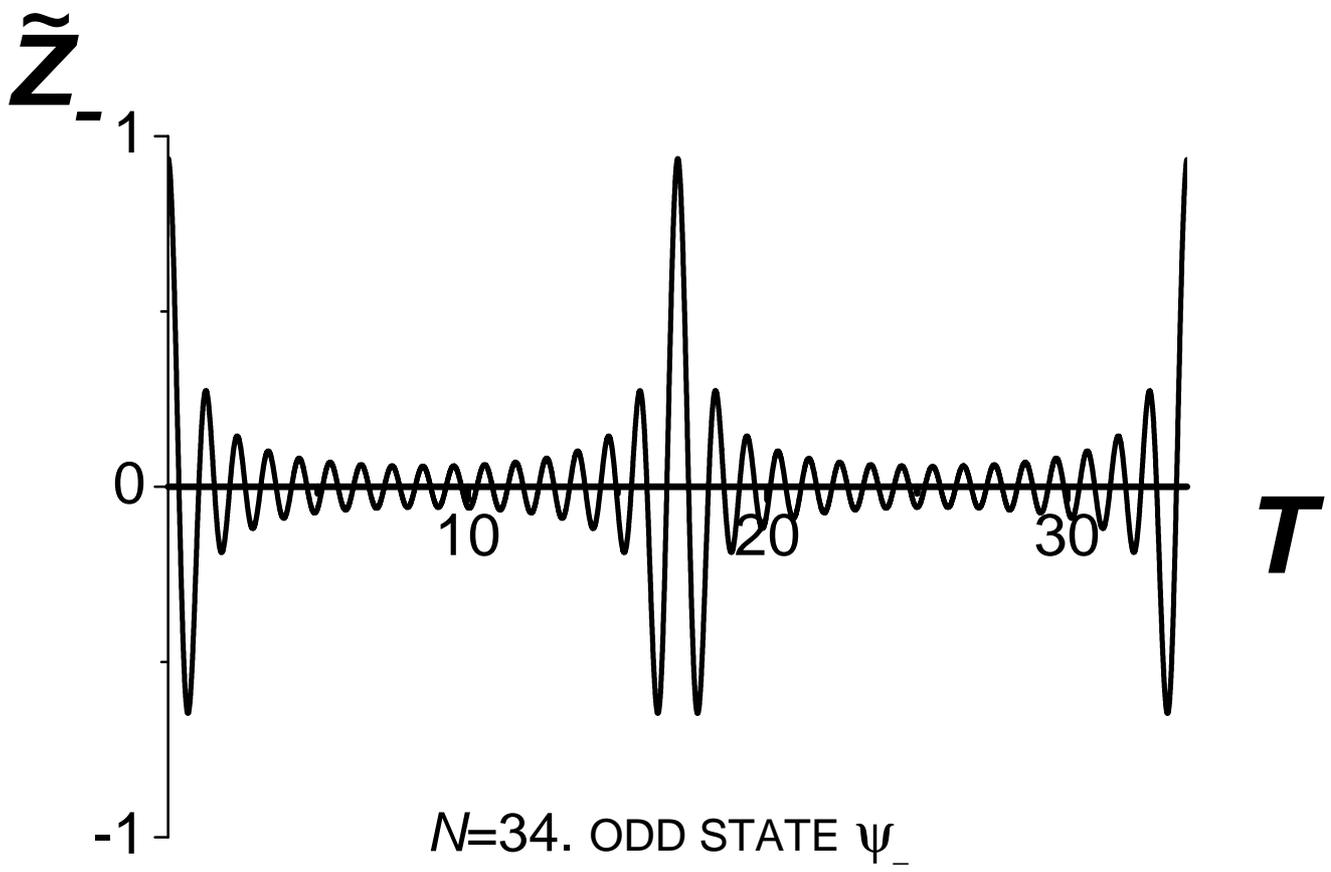
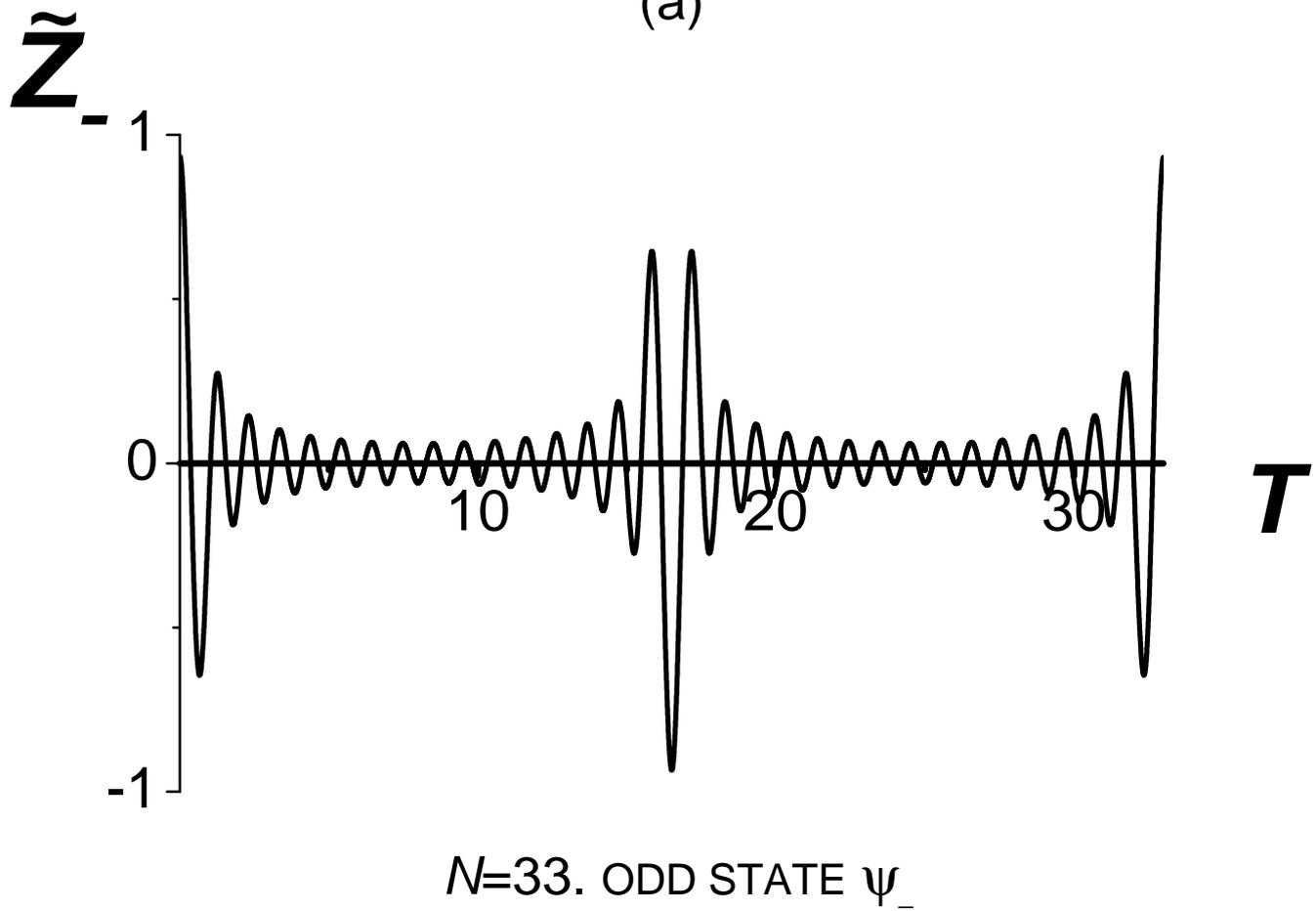
\end{document}